\documentclass[twocolumn,preprintnumbers,
amsmath,amssymb,aps,prd,nofootinbib,superscriptaddress]{revtex4}
\usepackage{graphicx}

\makeatletter
\renewcommand{\p@subsection}{}
\makeatother

\begin{document}

\title{Novel spectral broadening from
vector--axial-vector mixing in dense matter~\footnote{%
Talk given by M.~Harada at 
YITP workshop on 
``Thermal Quantum Field Theory and Their Applications 2009''
(September, 3 - 5, 2009, Yukawa Institute, Kyoto, Japan).
This talk is based on the work done in Ref.~\cite{VAmix-dense}.
}
}

\author{Masayasu Harada}
\affiliation{%
Department of Physics, Nagoya University,
Nagoya, 464-8602, Japan}
\author{Chihiro Sasaki}
\affiliation{%
Physik-Department,
Technische Universit\"{a}t M\"{u}nchen,
D-85747 Garching, Germany
}

\begin{abstract}
In this write-up we summarize main result of 
our recent analysis on the mixing
between transverse $\rho$ and $a_1$ mesons 
through a set of $\omega\rho a_1$-type interactions
in dense baryonic matter.
In the analysis, we showed that
a clear enhancement of the
vector spectral function appears below $\sqrt{s}=m_\rho$
for small three-momenta of the $\rho$ meson, 
and thus the vector spectrum exhibits broadening.
\end{abstract}

\maketitle

In-medium modifications of hadrons have been extensively
explored in the context of chiral dynamics of QCD~\cite{review,rapp}.
Due to an interaction with pions in the heat bath, the vector 
and axial-vector current correlators are mixed.
At low temperatures or densities a low-energy theorem based on 
chiral symmetry describes this mixing (V-A mixing)~\cite{theorem}.
The effects to the thermal vector spectral function have been
studied through the theorem~\cite{vamix}, or using chiral reduction
formulas based on a virial expansion~\cite{chreduction},
and near critical temperature in a chiral effective field theory 
involving the vector and axial-vector mesons as well as the 
pion~\cite{our}.

It has been derived, as a novel effect at finite baryon density,
that a Chern-Simons term leads to mixing between the vector and 
axial-vector fields in a holographic QCD model~\cite{hqcd}.
Unlike at zero density, the V-A mixing at finite density
appears in a tree-level Lagrangian.
This mixing modifies the dispersion relation of the transverse
polarizations and will affect the in-medium current correlation
functions independently of specific model dynamics.

In Ref.~\cite{VAmix-dense},
we focused on the V-A mixing at tree level and its consequence
on
the in-medium spectral functions which are the main input 
to the experimental observables.
We showed that 
the mixing produces
a clear enhancement of the vector spectral function 
below $\sqrt{s}=m_\rho$,
and that
the vector spectral function 
is broadened due to the mixing.
We also discussed its relevance to dilepton measurements.

At finite baryon density a system preserves parity 
but violates charge conjugation invariance.
Chiral Lagrangians thus in general build 
in the term
\begin{equation}
{\cal L}_{\rho a_1} = 2C\,\epsilon^{0\nu\lambda\sigma}
\mbox{tr}\left[ \partial_\nu V_\lambda \cdot A_\sigma
{}+ \partial_\nu A_\lambda \cdot V_\sigma \right]\,,
\label{vaterm}
\end{equation}
for the vector $V^\mu$ and axial-vector $A^\mu$ mesons with
the total anti-symmetric tensor $\epsilon^{0123}=1$ and a 
parameter $C$.
This mixing results in the dispersion relation~\cite{hqcd}
\begin{equation}
p_0^2 - \bar{p}^2 = \frac{1}{2}\left[ 
m_\rho^2 + m_{a_1}^2 \pm \sqrt{(m_{a_1}^2 - m_\rho^2)^2
{}+ 16 C^2 \bar{p}^2}
\right]\,,
\label{disp}
\end{equation}
which describes the propagation of a mixture of the transverse $\rho$ 
and $a_1$ mesons with non-vanishing three-momentum $|\vec{p}|=\bar{p}$.
The longitudinal polarizations, on the other hand,
follow the standard dispersion
relation, $p_0^2 - \bar{p}^2 = m_{\rho,a_1}^2$.
When the mixing vanishes as $\bar{p} \to 0$, Eq.~(\ref{disp}) with
lower sign provides $p_0 = m_\rho$ and it with upper sign
does $p_0 = m_{a_1}$.  In the following, we call the mode following
the dispersion relation with the lower sign in Eq.~(\ref{disp})
``the $\rho$ meson'', 
and that with the upper sign ``the $a_1$ meson''.

The mixing strength $C$ in Eq.~(\ref{vaterm}) can be 
estimated assuming the $\omega$-dominance
in the following way:
The gauged Wess-Zumino-Witten terms in 
an effective chiral Lagrangian include
the $\omega$-$\rho$-$a_1$ term~\cite{kaiser} which leads to
the following mixing term
\begin{equation}
{\cal L}_{\omega\rho a_1} = g_{\omega\rho a_1}
\langle \omega_0\rangle \epsilon^{0\nu\lambda\sigma}
\mbox{tr}\left[ \partial_\nu V_\lambda \cdot A_\sigma
{}+ \partial_\nu A_\lambda \cdot V_\sigma \right]\,,
\end{equation}
where the $\omega$ field is replaced with its expectation value
given by $\langle \omega_0\rangle = g_{\omega NN}\cdot n_B/m_\omega^2$.
One finds with empirical numbers
$C = g_{\omega\rho a_1}\langle \omega_0\rangle \simeq 0.1$\,GeV
at normal nuclear matter density. 
As we will show below, this is too small to have an importance
in the correlation functions.
In a holographic QCD approach, on the other hand,
the effects from an infinite tower of the $\omega$-type vector
mesons are summed up to give
$C \simeq 1\,\mbox{GeV}\cdot(n_B/n_0)$ 
with normal nuclear matter density $n_0 = 0.16$ fm$^{-3}$~\cite{hqcd}.
In the following we
assume an actual value of $C$ in QCD
in the range $0.1 < C < 1$\,GeV.
Some importance of the higher Kaluza-Klein
(KK) modes {\it even in vacuum} in the 
context of holographic QCD can be seen in the pion electromagnetic 
form factor at the photon on-shell:
This is saturated by the lowest four vector mesons in a top-down 
holographic QCD model~\cite{ss,matsuzaki2}. 
In hot and dense environment those higher members get modified
and the masses might be somewhat decreasing evidenced in an
in-medium holographic model~\cite{sstem}. This might provide
a strong V-A mixing $C > 0.1$\,GeV in three-color QCD
and the dilepton measurements may give a good testing ground.

\begin{figure}[htbp]
\begin{center}
\includegraphics[width=8cm]{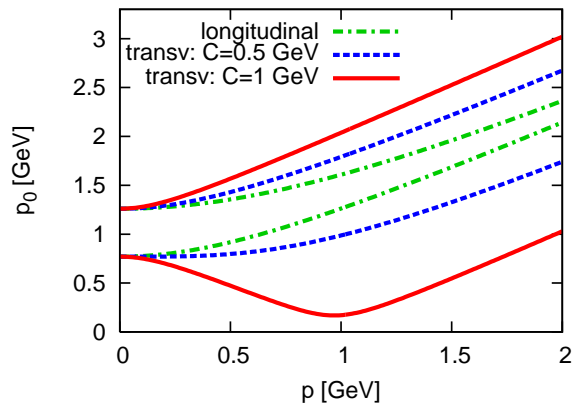}
\caption{
The dispersion relations for the $\rho$ (lower 3 curves)
and $a_1$ (upper 3 curves) mesons for $C=0.5$, and $1$\,GeV.
}
\label{dispersion}
\end{center}
\end{figure}

In Fig.~\ref{dispersion}, we show the dispersion 
relations~(\ref{disp}) for the transverse modes together
with those for the longitudinal modes with $C = 1$  and
$0.5$\,GeV.
This shows that, when $C=0.5$\,GeV, 
there are only small changes for both $\rho$ and $a_1$ mesons,
while a substantial change for $\rho$ meson when $C=1$\,GeV.
For very large $\bar{p}$ the longitudinal and transverse 
dispersions are in parallel with a finite gap, $\pm C$. 

\begin{figure*}
\begin{center}
\includegraphics[width=8.6cm]{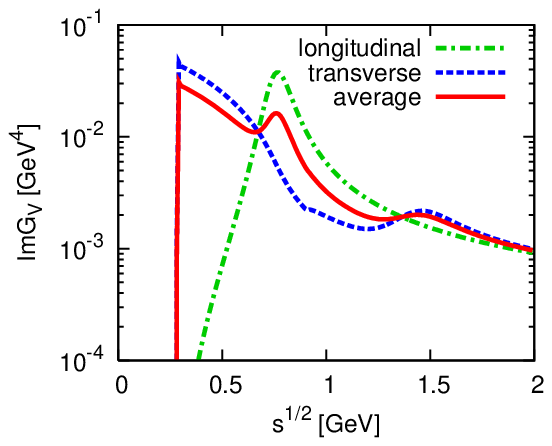}
\includegraphics[width=8.6cm]{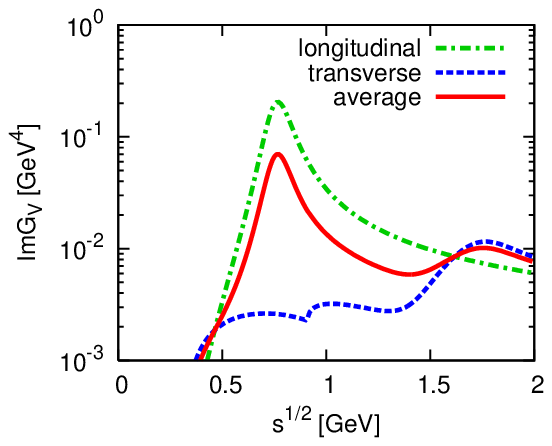}
\caption{
The vector spectral function for $C=1$\,GeV.
The curves of the left figure are calculated integrating over 
$0 < \bar{p} < 0.5$\,GeV, and those of the right figure
over $0.5 < \bar{p} < 1$\,GeV. 
Here we use the values of masses given by
$m_\pi = 0.14$\,GeV, $m_\rho = 0.77$\,GeV, 
$m_{a_1} = 1.26$\,GeV,
and the widths 
given by the imaginary 
part of one-loop diagrams in a chiral Lagrangian approach 
as~\cite{hs:ghls,VAmix-dense}
with the on-shell values of 
$\Gamma_{\rho}(s=m_\rho^2) = 0.15$\,GeV and
$\Gamma_{a_1}(s=m_{a_1}^2) = 0.33$\,GeV.
}
\label{CC1}
\end{center}
\end{figure*}

In Fig.~\ref{CC1},
we plot the integrated spectrum over three momentum,
which is a main ingredient in dilepton production rates.
Figure~\ref{CC1} (left) shows a clear enhancement of the spectrum 
below $\sqrt{s}=m_\rho$ due to the mixing.
This enhancement becomes much 
suppressed when the $\rho$ meson is moving with a large three-momentum
as shown in Fig.~\ref{CC1} (right). The upper
bump now emerges more remarkably and becomes a clear indication
of the in-medium effect from the $a_1$ via the mixing.

\begin{figure}[b]
\begin{center}
\includegraphics[width=8.6cm]{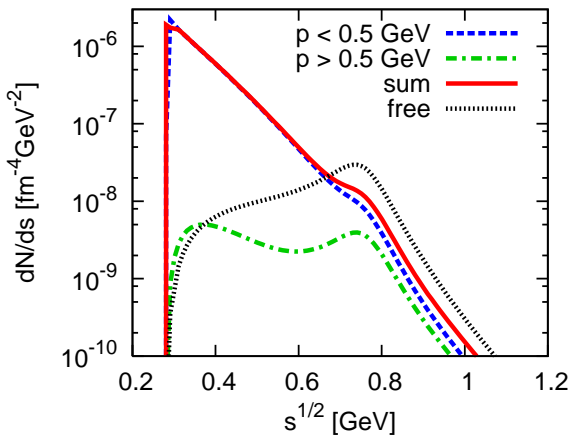}
\caption{
The dilepton production rate at $T=0.1$\,GeV for $C=1$\,GeV. 
Integration over $0 < \bar{p} < 0.5$\,GeV (dashed) and 
$0.5 < \bar{p} < 1$\,GeV (dashed-dotted) was carried out.
}
\label{rate}
\end{center}
\end{figure}

As an application of the above in-medium spectrum, we calculate
the production rate of a lepton pair emitted from dense 
matter through a decaying virtual photon.
Figure~\ref{rate} presents the integrated rate at $T=0.1$\,GeV
for $C=1$\,GeV.
One clearly observes a strong three-momentum dependence and an 
enhancement below $\sqrt{s}=m_\rho$ due to the Bose
distribution function which result in a strong spectral broadening.
The total rate is mostly governed by the spectrum with 
low momenta $\bar{p}<0.5$\,GeV due to the large mixing parameter $C$.
When density is decreased, 
the mixing effect gets irrelevant and
consequently in-medium effect in low $\sqrt{s}$ region is reduced 
in compared with that at higher density.  
The calculation performed in hadronic many-body theory in fact
shows that the $\rho$ spectral function with a low momentum
carries details of medium modifications~\cite{riek}.
One may have a chance to observe it in heavy-ion collisions 
with certain low-momentum binning at J-PARC, GSI/FAIR and RHIC 
low-energy running.

\begin{figure}[b]
\begin{center}
\includegraphics[width=8.6cm]{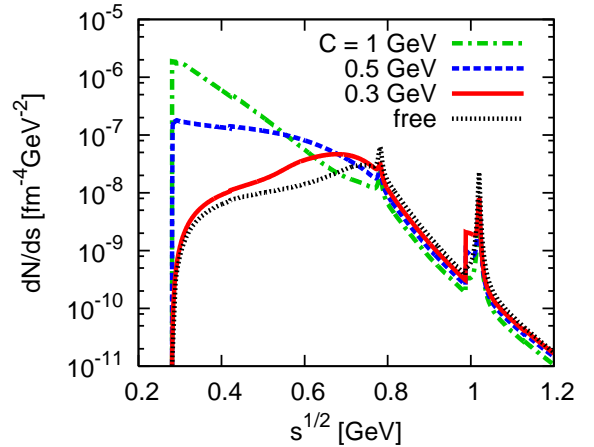}
\caption{
The dilepton production rate at $T=0.1$\,GeV
with various mixing strength $C$.
Integration over $0 < \bar{p} < 1$\,GeV was done.
We use the constant widths with values of
$\Gamma_\omega = 8.49$\,MeV, $\Gamma_\phi = 4.26$\,MeV,
$\Gamma_{f_1(1285)}=24.3$\,MeV and 
$\Gamma_{f_1(1420)}=54.9$\,MeV.
}
\label{rateV}
\end{center}
\end{figure}

It is straightforward to introduce other V-A mixing between 
$\omega$-$f_1(1285)$ and $\phi$-$f_1(1420)$.
In Fig.~\ref{rateV} we plot the integrated rate at $T=0.1$\,GeV
with several mixing strength $C$ which are phenomenological option.
One observes that the enhancement below $m_\rho$ is suppressed
with decreasing mixing strength. This forms into a broad bump
in low $\sqrt{s}$ region and its maximum moves toward $m_\rho$.
Similarly, some contributions are seen just below $m_\phi$.
This effect starts at threshold $\sqrt{s}= 2 m_K$. 
Self-consistent calculations of the spectrum in dense medium
will provide a smooth change and this eventually makes the $\phi$
meson peak somewhat broadened.

Finally, we remark that
the importance of the mixing effect studied here
relies on the coupling
strength $C$. 
Holographic QCD predicts an extremely strong 
mixing $C \sim 1$\,GeV at $n_B = n_0$ which leads to
vector meson condensation at $n_B \sim 1.1\,n_0$~\cite{hqcd}.
This may be excluded by known properties of nuclear
matter and therefore in reality the strength $C$ will be
smaller. We have discussed a possible range of $C$ to be
$0.1 < C < 1$\,GeV based on higher excitations and their in-medium 
modifications.
The parameter $C$ does carry an unknown density dependence.
This will be determined in an elaborated treatment of
hadronic matter along with the underlying QCD dynamics.
If $C \sim 0.1$\,GeV at $n_0$ were preferred as
the lowest-omega dominance, the mixing effect is irrelevant there.
However, it becomes more important at higher densities,
e.g. $C = 0.3$\,GeV at $n_B/n_0 = 3$ which leads to
a distinct modification from the spectrum in free space.

We would like to thank the organizers of the workshop
for giving an opportunity to present this work.
We acknowledge stimulating discussions with B.~Friman,
N.~Kaiser, S.~Matsuzaki, M.~Rho and W.~Weise.
The work of C.S. has been supported in part 
by the DFG cluster of excellence ``Origin and Structure of the 
Universe''.
The work of M.H. has been supported in part by
the JSPS Grant-in-Aid for Scientific Research (c) 20540262
and the Global COE Program of Nagoya University 
``Quest for Fundamental Principles in the Universe (QFPU)" 
from JSPS and MEXT of Japan.



\end{document}